# A Heteroscedastic Bayesian Generalized Logistic Regression Model with Application to Scaling Problems

Jack Sutton,[a*] Golnaz Shahtahmassebi,[b] Quentin S. Hanley[a,c] and Haroldo V. Ribeiro[d]

[a]*College of Science and Engineering, University of Derby, Markeaton Street, Derby DE22 3AW*
[b]*School of Science and Technology, Nottingham Trent University, Nottingham NG11 8NS*
[c]*GH and Q Services Limited, West Studios, Sheffield Road, Chesterfield S41 7LL, United Kingdom.*
[d]*Departamento de Física, Universidade Estadual de Maringá, Maringá, PR 87020-900, Brazil*

*Address for correspondence:* Jack Sutton, College of Science and Engineering, University of Derby, Markeaton Street, Derby, DE22 3AW, UK, Email: j.sutton@derby.ac.uk

**Abstract**

Power law scaling models have been used to understand the complexity of systems as diverse as cities, neurological activity, and rainfall and lightning. In the scaling framework, power laws and standard linear regression methods are widely used to estimate model parameters with assumed normality and fixed variance. Generalized linear models (GLM) can accommodate a wider range of distributions where the chosen distribution must meet the assumptions of the data to prevent model bias. We present a widely applicable Bayesian generalized logistic regression (BGLR) framework to more flexibly model a continuous real response addressing skew and heteroscedasticity. The Generalized Logistic Distribution (GLD) was selected to flexibly model skewed continuous data. This resulted in a nonlinear posterior distribution which may not have an analytical solution which can be solved numerically with Markov Chain Monte Carlo (MCMC) methods. We compared the BGLR model to standard and Bayesian normal models having fixed and varying variance when fitting power laws to 759 days of COVID-19 data. The BGLR yielded information beyond existing methods about the evolution of skew and skedasticity while revealing parameter bias of widely used methods. The BGLR flexibly modelled the complex characteristics necessary for an improved understanding of the propagation and dynamics of this infectious disease. The model is generally applicable and can be used as a template for modeling complexity with other distributions.

# 1 Introduction

Decades of research have explored the city effect on urban indicators. To encapsulate this relationship, a comprehensive generalization of allometric growth forming the foundations of urban scaling was presented [1]. This inspired further work in population [2–9] and density [10–13] scaling to model for a range of economic, health, crime, property and age indicators. This literature typically applies power laws (PLs) and standard linear regression techniques to estimate model parameters. Power law models have also been widely applied to problems as diverse as neurology [14], atmospheric science [15], and financial markets [16]. It is typically assumed that residual variance is fixed and normally distributed (i.e. $\varepsilon \sim N(0, \sigma^2)$). Previous work has shown that PL exponent estimates can vary depending on whether fixed or varying variance is assumed [17]. This work indicated that better models were needed to clarify the existence of nonlinear scaling. Furthermore, in other studies of scale, daily residual distributions of COVID-19 cases and deaths exhibited contracting and expanding variance along with positive and negative skew [12]. Current PL models lack support for these skew features as they are constructed based on Gaussian assumptions. Normality and fixed variance assumptions are inherited in the scaling field and continuing to rely on standard linear regression techniques will incorporate bias into all reported scaling model parameters. There is a need for a generalization allowing for varying variance and skewed distributions.

Standard linear regression models assume normally distributed, homoscedastic, linear and independent residuals [18–20]. For many data sets, these assumptions are too restrictive leading to bias in estimated model parameters. Generalized linear modelling (GLM) [21] provides a more flexible approach allowing for non-normal distributions. This generalization of ordinary linear regression provides a unifying framework for many commonly used statistical techniques. For instance, many of the statistical properties of the normal distribution are shared with a wider class of distributions known as the exponential family. A distribution belongs to the exponential family if it can be written in canonical form. In canonical form, it provides mathematically convenient canonical links that relate the mean of the response and the linear predictor. While using the canonical link is not mandatory, there are situations where the context might necessitate opting for a different link function [22].

The generalized logistic distribution (GLD) is suitable for modelling asymmetric continuous responses. The GLD is an extensively studied extension to the standard logistic distribution [23–25] and has a variety of forms [24,26–33]. The form considered here is the three-parameter (location, scale and shape) type I GLD [24,26–28]. It is currently unknown whether the type I GLD belongs to the exponential family. Nevertheless, the type I GLD shape parameter allows for skewed residuals allowing greater flexibility and, like the normal, it can support continuous data. The probability (PDF) and cumulative (CDF) density functions of the type I GLD are as follows:

$$f(x;\theta,\sigma,\alpha) = \frac{\alpha}{\sigma} \frac{\exp\left\{\frac{-(x-\theta)}{\sigma}\right\}}{\left[1+\exp\left\{\frac{-(x-\theta)}{\sigma}\right\}\right]^{\alpha+1}} \qquad (1)$$

and

$$F(x;\theta,\sigma,\alpha) = \frac{1}{\left[1+\exp\left\{\frac{-(x-\theta)}{\sigma}\right\}\right]^{\alpha}} \qquad (2)$$

where $\theta$, $\sigma$ and $\alpha$ are the location, scale and shape parameters respectively. $\theta \in \mathbb{R}$, $\alpha > 0$, $\sigma > 0$ and $-\infty < x < +\infty$. The shape parameter $\alpha$ indicates the type of skewness. When $\alpha > 1$, the distribution is positively skewed, when $\alpha = 1$ the distribution is symmetrical and when $\alpha < 1$ the distribution is negatively skewed. The following parameterisation $f(x; \theta, \sigma, \alpha) = f(x; 0,1,1)$ reduces the PDF (Equation (1)) and CDF (Equation (2)) to the standard logistic distribution [23–25] such that Equation (1) reduces to $f(x; 0,1,1) = \exp(-x)\{1 + \exp(-x)\}^{-2}$.

A proof [28], sec. 3) has been published showing that maximum likelihood estimates (MLE) for the type I GLD do not exist. However, more recent work [27] claimed this result was due to unintended logarithm and summation errors. Order statistics have been extensively studied for $\alpha$ for the Type I GLD [25] and tables corresponding to the means, variances and covariances are available. Using these tables, the study derived the best linear unbiased estimators of the location and scale parameters assuming a known shape parameter and presented them in tabulated format. The type I GLD is readily accessible with sampling and fitting tools [34].

The GLD has been applied to a range of problems. For example, in a study of inflation rates, the GLD captured changes in mean, variance and skewness [35]. More recently, the GLD was used to study residuals following regression analysis with assumed normality of COVID-19 cases and deaths over a 15-month period across England and Wales. The residual distributions were better characterized using the GLD density curve than the normal density curve illustrating the flexibility of the GLD [13]. However, since the GLD model was applied after regression methods (with assumed normality) due to a lack of a unified GLD-based framework, there is a need to introduce one.

We present a Bayesian generalized logistic (GL) regression model able to treat heteroscedastic and skewed (both positive and negative) continuous data. First (Section 2), we present the statistical properties and graphical representations of the GLD. Second (Section 3), the Bayesian generalized logistic (GL) regression model is developed along with a complete framework using Markov chain Monte Carlo (MCMC) methods (Section 4). Third (Section 5), we compare the GL regression model to normal methods using PLs and data from the COVID-19 pandemic. Finally, concluding remarks are given (Section 6) along with suggestions for further generalization and application.

## 2 Properties of the type I Generalized Logistic Distribution

### 2.1 GLD Maximum Likelihood Estimation

A MLE for the generalized logistic has been derived [27]. Briefly, let $X_1, X_{2,\ldots,}X_n$ be independent random variables taken from the continuous GLD given by $f(x; \theta, \sigma, \alpha)$. Henceforth, the PDF corresponding to the GLD will be referred to as $GLD(x; \theta, \sigma, \alpha)$. Suppose that a sample of observations $X_1, X_{2,\ldots,}X_n$ is taking the values $x_1, x_{2,\ldots,}x_n$ then the likelihood is $L(\theta, \sigma, \alpha | x_1, x_{2,\ldots,}x_n) = GLD(x_1, x_{2,\ldots,}x_n; \theta, \sigma, \alpha) = GLD(x_1; \theta, \sigma, \alpha) \times GLD(x_2; \theta, \sigma, \alpha) \times \ldots \times GLD(x_n; \theta, \sigma, \alpha) = \prod_{i=1}^{n} GLD(x_i; \theta, \sigma, \alpha)$. The likelihood function for the GLD is given by:

$$L = L(\theta, \sigma, \alpha | x_1, x_{2,\ldots,}x_n) = \prod_{i=1}^{n} \frac{\alpha}{\sigma} \frac{\exp\left\{\frac{-(x_i-\theta)}{\sigma}\right\}}{\left[1+\exp\left\{\frac{-(x_i-\theta)}{\sigma}\right\}\right]^{\alpha+1}} \tag{3}$$

Thus, the corresponding GLD log-likelihood function $\log\{L(\theta, \sigma, \alpha|x_i)\}$ is given by:

$$\log(L) = \log\{L(\theta, \sigma, \alpha|x_i)\} = n\log\left(\frac{\alpha}{\sigma}\right) - \sum_{i=1}^{n}\left(\frac{x_i-\theta}{\sigma}\right) - (\alpha+1)\sum_{i=1}^{n}\log\left[1+\exp\left\{\frac{-(x_i-\theta)}{\sigma}\right\}\right] \tag{4}$$

To obtain the MLE, replace $(\theta, \sigma, \alpha)$ by its estimators $(\hat{\theta}, \hat{\sigma}, \hat{\alpha})$ and find values that jointly maximize the likelihood function. For example, it has been shown [27] that setting $\frac{\partial \log (L)}{\partial \alpha} = 0$ gives the following:

$$\hat{\alpha} = \frac{n}{\sum_{i=1}^{n} \log\left[1+\exp\left\{\frac{-(x_i-\hat{\theta})}{\hat{\sigma}}\right\}\right]} \tag{5}$$

where it can be seen that $\hat{\alpha} \to \infty$ if $\hat{\theta} \to -\infty$ and that $\hat{\alpha} \to 0$ if $\hat{\sigma} \to 0$. Generally, the MLEs $(\hat{\theta}, \hat{\sigma}, \hat{\alpha})$ can be obtained if there exist $\hat{\theta} \in \mathbb{R}, \hat{\sigma} > 0$ and $\hat{\alpha} > 0$ which simultaneously maximise the log likelihood function $\log(L(\theta, \sigma, \alpha | x_i))$ given in Equation (4). This can be done by applying numerical methods such as the Newton-Raphson procedure which have been shown to collapse if $\hat{\alpha} = \infty$ and $\hat{\theta} = -\infty$ or when $\hat{\alpha} = 0$ and $\hat{\sigma} = 0$ [27]. If this happens then it indicates that the GLD is not valid for the data set being modelled. Collapse can indicate that other distributions such as the Gumbel or two-parameter reciprocal exponential are more appropriate [27].

## 2.2 Method of Moments

Estimators, $(\hat{\theta}, \hat{\sigma}, \hat{\alpha})$, can be derived for the parameters $(\theta, \sigma, \alpha)$ that are asymptotically unbiased and consistent [28]. These estimators are functions of the first three sample moments corresponding to mean, variance and skew as follows:

$$E(X) = \theta + \sigma\{\psi(\alpha) - \psi(1)\} \tag{6}$$

$$\text{Var}(X) = \sigma^2\{\psi'(1) + \psi'(\alpha)\}$$

$$\text{Skew}(X) = \frac{\psi''(\alpha) - \psi''(1)}{\{\psi'(\alpha) + \psi'(1)\}^{\frac{3}{2}}}$$

where $\psi(\cdot)$ is the digamma function, and $\psi'(\cdot)$ and $\psi''(\cdot)$ are its first and second derivatives, respectively.

## 2.3 GLD Graphical Presentation

Suitably parameterized, the GLD is a flexible and robust distribution for modelling continuous data. The GLD reduces to the standard logistic distribution when $\theta = 0$ and $\sigma = \alpha = 1$ (Figure 1(a)). Continuing with a fixed $\theta = 0$ and $\alpha = 1$ and changing $\sigma$ retains a symmetrical shape. Increasing $\sigma$ widens the span over $X$ (Figure 1(b)). The shape parameter indicates the skew of the distribution. If $\alpha < 1$, the distribution is negatively skewed (Figure 2(a)) and positive skew is indicated $\alpha > 1$ (Figure 2(b)). In each case (negative and positive skew) the tails of the distribution become heavier with increasing $\alpha$. Varying $(\theta, \sigma, \alpha)$ allows the GLD to characterize a range of different complex shapes (Figure 3).

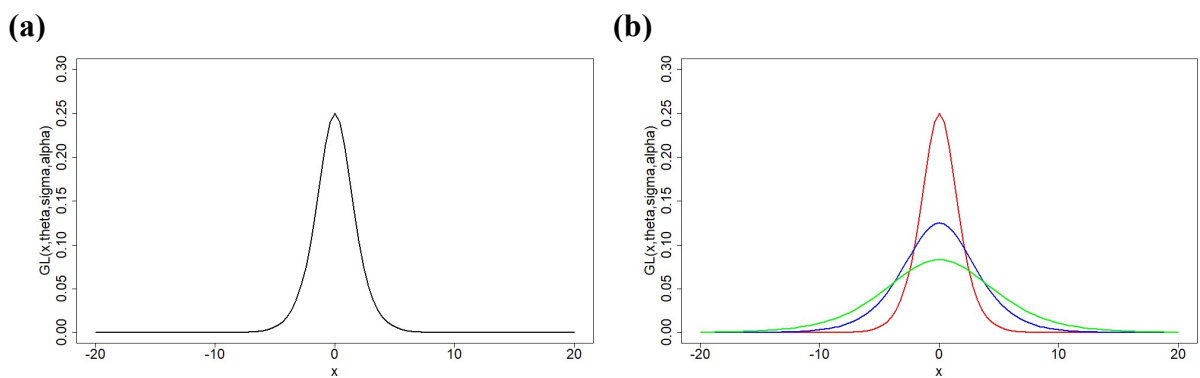

**Figure 1.** The standard logistic distribution (a) and the type I GLD (b). The solid black line in panel (a) represents the standard logistic distribution with the following parameterization: $\theta = 0, \sigma = \alpha = 1$. In panel (b) the solid green, blue and red density curves have the following parameterizations: $f(x; \theta, \sigma, \alpha) = f(x; 0, 3, 1)$, $f(x; \theta, \sigma, \alpha) = f(x; 0, 2, 1)$ and $f(x; \theta, \sigma, \alpha) = f(x; 0, 1, 1)$ respectively.

(a)  (b)

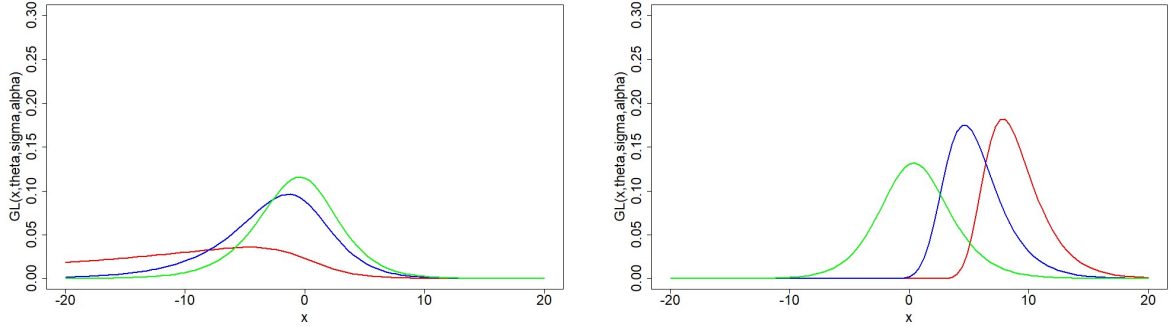

**Figure 2.** The GLD with location and scale parameters set to $\theta = 0$ and $\sigma = 2$ resprectively. Panel (a) illustrates negative skew where the shape parameter is set to $\alpha = 0.8$ (solid green line), $\alpha = 0.5$ (solid blue line) and $\alpha = 0.1$ (solid red line). While panel (b) illustrates positive skew where the shape parameter is set to $\alpha = 1.2$ (solid green line), $\alpha = 10$ (solid blue line) and $\alpha = 50$ (solid red line).

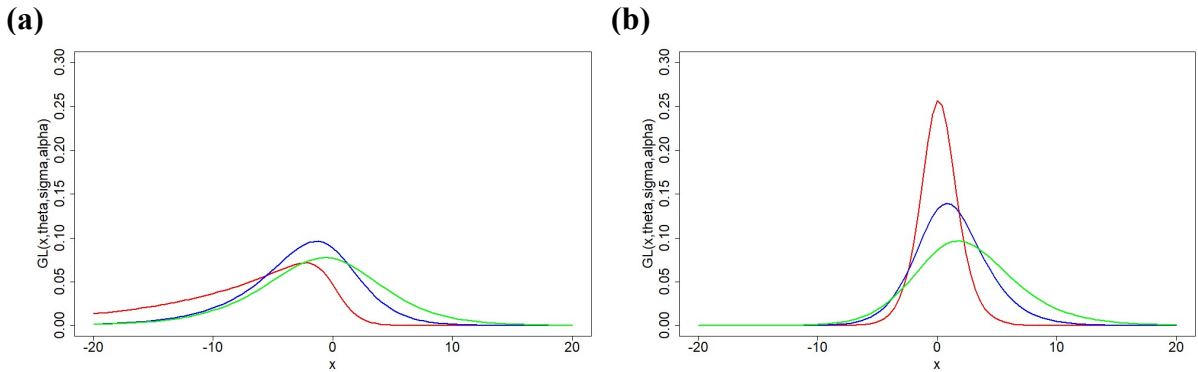

**Figure 3.** The GLD with location values set $\theta = 0$ illustrating negative (panel a) and positive (panel b) skew exhibiting a heavier left tail with increasing $\alpha$. In panel (a) scale and shape parameters are set to $\sigma = 3$, $\alpha = 0.8$ (solid green line), $\sigma = 2$, $\alpha = 0.5$ (solid blue line) and $\sigma = 1$, $\alpha = 0.1$ (solid red line). While in panel (b) scale and shape parameters are set to $\sigma = 3$, $\alpha = 1.8$ (solid green line), $\sigma = 2$, $\alpha = 1.5$ (solid blue line) and $\sigma = 1$, $\alpha = 1.1$ (solid red line). Varying $\sigma$ also allows for a variety of other complex shapes further showcasing the flexibility of the GLD.

The GLD does not collapse to a normal, however, based on simulations we found it to be a good approximation (Figure 4). The simulations consisted of 4 sets of 10,000 normally distributed random numbers with a mean $\mu_{\text{normal}} = 0$ and varying standard deviation (SD) ($\sigma_{\text{normal}} = 1, 3, 5, 7$) corresponding to the normal distribution. The shape parameter $\alpha$ in the GLD was near 1 in all simulations indicating no skew. The GLD fits to low $\sigma$ tended to have heavier tails and an elevated peak compared to fitting a normal density curve. Tails of the GLD and normal become more alike with increasing $\sigma$ although the slightly elevated peak remains.

Although the GLD does not exactly collapse to the normal distribution, the capacity to model both symmetrical and asymmetrical distributions is an advantageous feature allowing for skew.

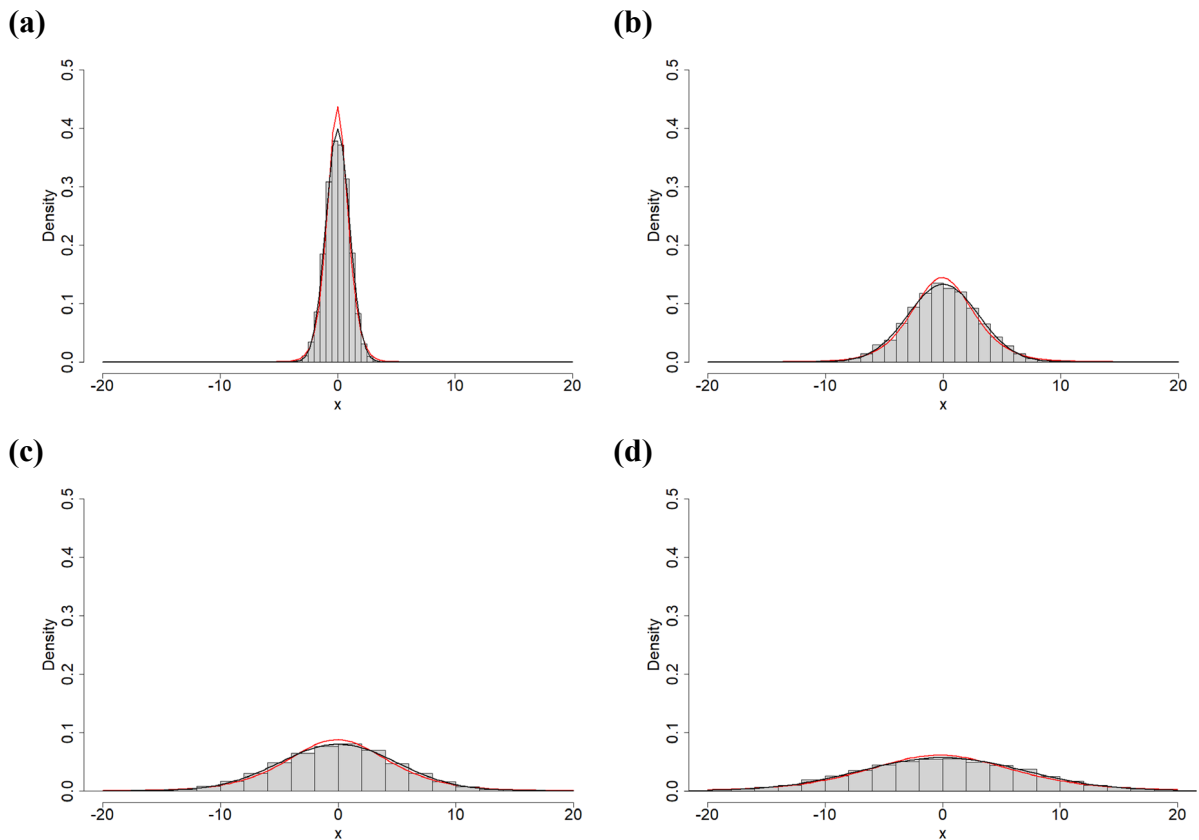

**Figure 4.** Four simulations of 10,000 randomly normally distributed data with fitted normal (solid black line) and GL (solid red line) density curves. The simulated data all have a mean $\mu_{normal} = 0$ with standard deviation of (a) $\sigma_{normal} = 1$, (b) $\sigma_{normal} = 3$, (c) $\sigma_{normal} = 5$ and (d) $\sigma_{normal} = 7$.

## 3 The GL Regression Model (GLR)

Consider $Y'_1, \ldots, Y'_n$ an $n$ sample of continuous data following a GLD with PDF and CDF in the form of Equations (1) and (2), respectively, such that $Y'_i \sim GLD(x_i; \theta, \sigma, \alpha)$ for $i = 1, \ldots, n$ where $\theta$ is the location, $\sigma$ is the scale and $\alpha$ is the shape parameters. We model the mean and variance using the following link functions:

$$E(Y_i') = g_1(\theta) = x_i^T \boldsymbol{\beta} = \eta_i \tag{7}$$

$$\log\{\text{Var}(Y_i')\} = g_2(\sigma) = x_i^T \boldsymbol{\beta}' = \eta_i'$$

where $g_1(\cdot)$ and $g_2(\cdot)$ denote monotone, twice differentiable link functions, $x_i$ is the $i$th column of a design matrix of size $p \times n$, where $p$ is the number of explanatory variables, $\boldsymbol{\beta}$ is $p \times 1$ vector of unknown mean regression coefficients and $\boldsymbol{\beta}'$ is a $p \times 1$ vector of unknown variance regression coefficients. The mean linear predictor denoted as $\eta_i$ relates to the link function $g_1(\cdot)$ such that $\theta_i = g_1^{-1}(\eta_i) = g_1^{-1}(x_i^T \boldsymbol{\beta})$ and the variance linear predictor denoted as $\eta_i'$ relates to the link function $g_2(\cdot)$ such that $\sigma_i = g_2^{-1}(\eta_i') = g_2^{-1}(x_i^T \boldsymbol{\beta}')$. In the model, $g_1(\cdot)$ is the identity link function and $g_2(\cdot)$ is the logarithmic link function.

The log-likelihood function corresponding to the GLD, given in Equation (4), is adjusted to account for the regression model proposed in Equation (7) for the mean $E(Y_i')$ and variance $\log(\text{Var}(Y_i'))$ functions. Taking this into consideration, the log-likelihood given in Equation (4) is adjusted to account for the proposed link functions given in Equation 7 such that:

$$\log(L) = \log\{L(\boldsymbol{\beta}, \boldsymbol{\beta}', \alpha | x_i, y_i)\} = n \log\left(\frac{\alpha}{x_i^T \boldsymbol{\beta}'}\right) - \sum_{i=1}^{n} \left(\frac{y_i - x_i^T \boldsymbol{\beta}}{x_i^T \boldsymbol{\beta}'}\right) - (\alpha + 1) \sum_{i=1}^{n} \log\left[1 + \exp\left\{\frac{-(y_i - x_i^T \boldsymbol{\beta})}{x_i^T \boldsymbol{\beta}'}\right\}\right] \tag{8}$$

where the location, $\theta$, and scale, $\sigma$, parameters are replaced with the proposed linear predictors. This results in the log-likelihood corresponding to our developed GL regression model. The GL regression log-likelihood (Equation (8)) will be applied in our Bayesian framework.

# 4 Bayesian Generalized Logistic Regression (BGLR)

## 4.1 Prior Distribution

For the Bayesian framework, we specify prior distributions for each model parameter. Since no information is provided, non-informative priors are assigned to the mean ($\boldsymbol{\beta}$) and variance ($\boldsymbol{\beta'}$) for each regression coefficient along with the GLD shape parameter $\alpha$. To express prior ignorance, each $\boldsymbol{\beta}$ and $\boldsymbol{\beta'}$ is assumed normally distributed with a mean of 0 and large variance ($10^4$) such that:

$$
\begin{aligned}
\beta_0 &\sim N(0, \sigma_{\beta_0}^2) & \beta_0' &\sim N(0, \sigma_{\beta_0'}^2) \\
\beta_1 &\sim N(0, \sigma_{\beta_1}^2) & \beta_1' &\sim N(0, \sigma_{\beta_1'}^2) \\
&\vdots & &\vdots \\
\beta_p &\sim N(0, \sigma_{\beta_p}^2) & \beta_p' &\sim N(0, \sigma_{\beta_p'}^2)
\end{aligned}
\quad (9)
$$

where $\sigma_{\beta_p}^2$ is some large variance for $p = 0, 1, \ldots, P$. Furthermore, a non-informative gamma distribution is considered for $\alpha$ such that $\alpha \sim \Gamma(a_1, b_1)$ where $a_1 = b_1 = 1$. The Bayesian approach is advantageous since priors can be updated when external information is available or using elicitation techniques [36]. It is also possible to set them up sequentially using the previous posterior distribution as the current prior distribution [36].

## 4.2 Posterior Distribution

In a Bayesian approach, statistical inference is performed on the posterior distribution of the model parameter vector, $\psi$. In the BGLR model, we consider the parameter vector given by:

$$\psi = (y'_i; \boldsymbol{\beta}, \boldsymbol{\beta}', \alpha) \tag{10}$$

where $\boldsymbol{\beta} = (\beta_0, \beta_1, \ldots, \beta_p)$ is the set of mean regression coefficients and $\boldsymbol{\beta}' = (\beta'_0, \beta'_1, \ldots, \beta'_p)$ is the set of variance regression coefficients. Given the priors and likelihood function, the joint posterior density function can be expressed as:

$$f_{\text{post}}(\psi|y') \propto f_{\text{like}}(y'|\psi) f_{\text{prior}}(\psi) \tag{11}$$

where $f_{\text{like}}(y'|\psi)$ is given by:

$$f_{\text{like}}(y'|\psi) = \prod_{i=1}^{n} GLD(y'_i; \boldsymbol{\beta}, \boldsymbol{\beta}', \alpha) \tag{12}$$

and $f_{\text{prior}}(\psi)$ is completed with independent prior distributions, given by:

$$f_{\text{prior}}(\psi) = f(\beta_0) \cdot f(\beta_1) \cdot \cdots \cdot f(\beta_p) \cdot f(\beta'_0) \cdot f(\beta'_1) \cdot \cdots \cdot f(\beta'_p) \cdot f(\alpha) \tag{13}$$

We were unable to confirm whether the posterior distribution is analytically tractable which is often the case for nonlinear and non-Gaussian models. Thus, to generate samples from the posterior distribution, we used MCMC methods by applying the random walk Metropolis-Hasting (MH) algorithm. Inference on the parameter vector, $\psi$, can be based on the posterior summaries of the marginal posterior distribution such as the mean, median, standard deviation and quantiles.

## 4.3 Metropolis-Hasting Algorithm

MCMC methods are a broad set of algorithms used to sample a probability distribution [36–39]. In the absence of an analytical solution to the GLD posterior distribution, the BGLR applies MCMC methods [36,40,41] to sample from the posterior distribution. MCMC methods sequentially sample from a simple candidate distribution where the next sample value depends on the current sample value and is either accepted or rejected based on some probability. If the sample is accepted, then the candidate value is updated and applied to the next iteration while, if rejected, the current sample is continued into the next iteration. After $m$ iterations the sample values approach the desired target distribution. Appropriate candidate distributions must: (a) have the same state space; (b) have sample draws converge to the target distribution; and (c) easily accessible random draws. If the candidate distribution meets these requirements in the context of the BGLR model, the random walk MH algorithm proceeds as shown in Algorithm 1 [36,41,42].

**Algorithm 1** Metropolis-Hastings Algorithm

---

(a) Draw an initial estimated value in the parameter vector $\psi^0$ from the candidate distribution.

(b) Repeat for $t = 1, \dots, m$ where $m$ is the number of MCMC chains and

   (i) Draw a sample value from the candidate distribution such that

   $$\psi^{\text{cand}} \sim q(.\,|\psi^{t-1})$$

   (ii) Calculate the acceptance ratio $\lambda$, used to either accept or reject the sample candidate value. The acceptance is obtained by applying the following:

   $$\lambda = \frac{f_{\text{post}}(\psi^{\text{cand}}|y')q(\psi^{(t-1)}|\psi^{\text{cand}})}{f_{\text{post}}(\psi^{(t-1)}|y')q(\psi^{\text{cand}}|\psi^{(t-1)})}$$

---

(iii) Set

$$\psi^t = \begin{cases} \psi^{\text{cand}} & \text{with probability min } (\lambda, 1) \\ \psi^{(t-1)} & \text{otherwise} \end{cases}$$

(c) Increment the iteration: $t = t + 1$ and repeat step 2 until $m$ chains have been implemented.

To implement the MH algorithm above, the mean $\boldsymbol{\beta}$ and variance $\boldsymbol{\beta}'$ parameters were assigned normal (as described in Section 4.1) given by:

$$\boldsymbol{\beta}_p \sim N\left\{\mu_{\text{normal}}^{(t-1)}, \sigma_{\text{normal}}^2\right\} \tag{14}$$

$$\boldsymbol{\beta}'_p \sim N\left\{\mu_{\text{normal}}^{(t-1)}, \sigma_{\text{normal}}^2\right\} \tag{15}$$

where $p = 0, 1, \ldots, P$ are the of number of parameters in each vector space $\boldsymbol{\beta}$ and $\boldsymbol{\beta}'$. For a normal proposal, the mean is centered at the value of the previous iteration with some suitably chosen variance. We chose not to use a similar formalism for the shape parameter, $\alpha$, as it exclusively maps to the domain of positive real numbers. Instead, $\alpha$ was chosen to be a gamma given by:

$$\alpha \sim \Gamma(a^{(t-1)}, b^{(t-1)}) \tag{16}$$

where $a^{(t-1)} = \frac{(\mu_{\text{gamma}}^{t-1})^2}{(\sigma_{\text{gamma}}^{t-1})}$ and $b^{(t-1)} = \frac{(\mu_{\text{gamma}}^{t-1})}{(\sigma_{\text{gamma}}^{t-1})}$ for $t = 1, \ldots, m$ where $m$ is the total number of MCMC chains.

*4.4 DIC Score*

To compare the model performance between Bayesian models, a deviance information criterion (DIC) [43] score was computed to measure the goodness-of-fit. DIC is somewhat a Bayesian version of the Akaike information criterion (AIC), where the posterior distribution has been obtained using the MCMC chains. As illustrated [36], the DIC deals with MCMC methods as follows:

$$\text{computed } f_{\text{DIC}} = 2 \left\{ \log f_{\text{like}}(y'|\hat{\psi}) - \frac{1}{m} \sum_{t=1}^{m} \log f_{\text{like}}(y'|\psi^t) \right\} \tag{17}$$

where $f_{\text{like}}(y'|\psi)$ is the GLD likelihood function and $\psi$ is the parameter vector. The second term is the average of $\psi$ over its posterior distribution across all calculated MCMC chains where $m$ is the total number of iterations. The DIC provides a trade-off between the goodness of fit and the complexity of the model. The actual DIC, using the computed $f_{\text{DIC}}$ is defined in terms of the deviance rather than the log predictive density, such that:

$$\text{DIC} = -2\log f(y'|\hat{\psi}_{\text{Bayes}}) + 2f_{\text{DIC}} \tag{18}$$

where a lower DIC indicates a better balance between model fit and complexity, making it a useful tool for model selection in Bayesian analysis.

*4.5 R Packages, Data, and Implementation*

The COVID-19 and population data were analyzed using R version (4.1.2) [44]. Population data from the 2011 census and regional land areas were obtained from NOMIS (https://www.nomisweb.co.uk). The COVID-19 data were obtained from the UK Coronavirus dashboard (https://coronavirus.data.gov.uk/) currently maintained by the UK Health Security Agency. The API was used to collect the data which was formatted with the rio (0.5.27) [45], xlsx (0.5.7) [46], httr (1.4.2) [47], plyr (1.8.5) [48], and lubridate (1.7.9.2) [49] packages. The tools for the type I generalized logistic distribution we acquired using the glogis (1.0–1) [34] package. Matrix statistics to obtain posterior summaries were computed using the matrixStats (0.61.0) [50] package. Gelman and Rubin's convergence diagnostics were performed using the coda (0.19-4) [51] package. The R-code is available as supplementary information (online Appendix A).

## 5 Example

To illustrate the BGLR model, we consider the English and Welsh COVID-19 and population density data from 337 lower tier local authorities over a 759 day period beginning 01/03/2020 and ending 29/03/2022 using a (PL) scaling model. This timeframe included different testing regimes and a range of government interventions (lockdowns, restrictions, vaccine programme, etc.). A snapshot of the formatted data set is given in Table 1 and the complete data set is provided as supplementary information (online Appendix B). Issues related to the data set and formatting it for analysis have been discussed previously [13]. This analysis adds 285 days to a previous analysis done using a simpler model [13]. To illustrate the advantages of the BGLR, we compare it to previous normality methods and assess bias when assuming a normal system.

**Table 1.** A representation of the COVID-19 data set. The dataset consists of lower tier local authority (LTLA) regions aligned to population, area and daily COVID-19 case statistics.

| Region | Population | Area (Hectares) | Day 1 | Day 2 | Day 3 | ⋯ | Day 759 |
|---|---|---|---|---|---|---|---|
| Adur | 61167 | 4180.71 | 0 | 0 | 0 | ⋯ | 33 |
| Allerdale | 96468 | 124158.29 | 0 | 0 | 0 | ⋯ | 69 |
| ⋮ | ⋮ | ⋮ | ⋮ | ⋮ | ⋮ | ⋱ | ⋮ |
| York | 197808 | 27193.63 | 0 | 1 | 0 | ⋯ | 147 |

Here we apply the population density PL model along with its linearized version are as follows:

$$Y_D = Y_0 P_D^\beta 10^\varepsilon \qquad (19)$$

$$\log(Y_D) = \log(Y_0) + \beta \log(P_D) + \varepsilon$$

where $Y_D = Y/A$ is the indicator (COVID-19 cases) density, $P_D = P/A$ is the population density where $A$ is the area of a region, $Y_0$ is the pre-exponential factor, $\beta$ is the PL exponent and $\varepsilon$ are residuals that are independent and identically distributed with common $N(0, \sigma^2)$. Rewriting Equation (19) into standard regression form gives:

$$Y_i = \beta_0 + \beta_1 x_i + \varepsilon_i \qquad (20)$$

where $Y_i = \log(Y_{D_i})$, $\beta_0 = \log(Y_0)$, $\beta_1 = \beta$, $x_i = \log(P_{D_i})$ and $\varepsilon_i = \varepsilon$. The BGLR framework allows for asymmetric and heteroscedastic data sets as follows:

$$E(Y_i') = g_1(\theta) = x_i^T \boldsymbol{\beta} = \beta_0 + \beta_1 x_i = \eta_i \qquad (21)$$

$$\log\{\text{Var}(Y_i')\} = g_2(\sigma) = x_i^T \boldsymbol{\beta}' = \beta_0' + \beta_1' x_i = \eta_i'$$

where $Y_i' \sim GLD(x_i; \theta, \sigma, \alpha)$. The parameters $\beta_0$ and $\beta_1$ are PL regression coefficients with variance coefficients $\beta_0'$ and $\beta_1'$. In this example, $\beta_0'$ corresponds to traditional variance, while $\beta_1'$ is a scedasticity parameter such that: $\beta_1' = 0$ for homoscedastic; $\beta_1' < 0$ for negatively heteroscedastic (decreasing with $x_i$); and $\beta_1' > 0$ for positively heteroscedastic (increasing with $x_i$). A straight-line model of variance has been used here; however, this is not a requirement allowing more flexible models and more detailed metrics of scedasticity

*5.1 MCMC Results*

Gelman and Rubin's convergence diagnostics were used [42,52] with values less than 1.1 indicating convergence and values greater than 1.1 indicating non-stationary chains. The MH algorithm was implemented with $m = 20{,}000$ iterations with the first 10,000 draws disregarded (burn-in). In this way, we found most estimates for $\beta_0$, $\beta_1$, $\theta$, $\sigma$ and $\alpha$ with recognized convergence over 759 days and in some cases estimates for $\beta_0'$ and $\beta_1'$ (online Appendix C). Despite the existence of potentially more accurate estimates for $\beta_0'$ and $\beta_1'$, the BGLR model still provides a flexible alternative and in many cases outperforms normal methods. We note that heteroscedasticity was not apparent on all days.

*5.2 Model Parameters*

Where possible, we compared our BGLR model (Figures 5-7, black) to the standard linear regression (SLR) model with fixed variance (Figures 5-7, red) and to a Bayesian normal regression (BNR) model with varying variance (Figures 5-7, blue). Computed differences in parameter estimates between BGLR and BNR models are also available (Figures 5-7). Positive differences indicate increased estimates and negative differences indicate decreased estimates in the BGLR model. The BGLR model obtains 5 parameters: 2 for $\theta$, $(\hat{\beta}_0, \hat{\beta}_1)$, representing the PL (Figure 5); 2 for $\sigma$, $(\beta_0', \beta_1')$, the PL variance parameters (Figure 6), and one for $\alpha$ (Figure 7).

Although daily pre-exponential factors $\hat{\beta}_0$ (Figure 5(a)) and exponents, $\hat{\beta}_1$ (Figure 5(c)) generally track each other, computed differences highlight structured bias in the estimates (Figure 5; Panels (b) and (d)) for the BNR. The BGLR posterior mean and corresponding 95% credible intervals of the posterior distributions of the model parameters are represented in grey. The most noticeable differences and wider computed CIs occurred at the very beginning of the pandemic, summer 2020 and 2021 when fewer regions reporting cases and the number of cases was lower. Some of the apparent instability in the $\hat{\beta}_0$ parameter estimates is due to trends in the weekly reporting cycle.

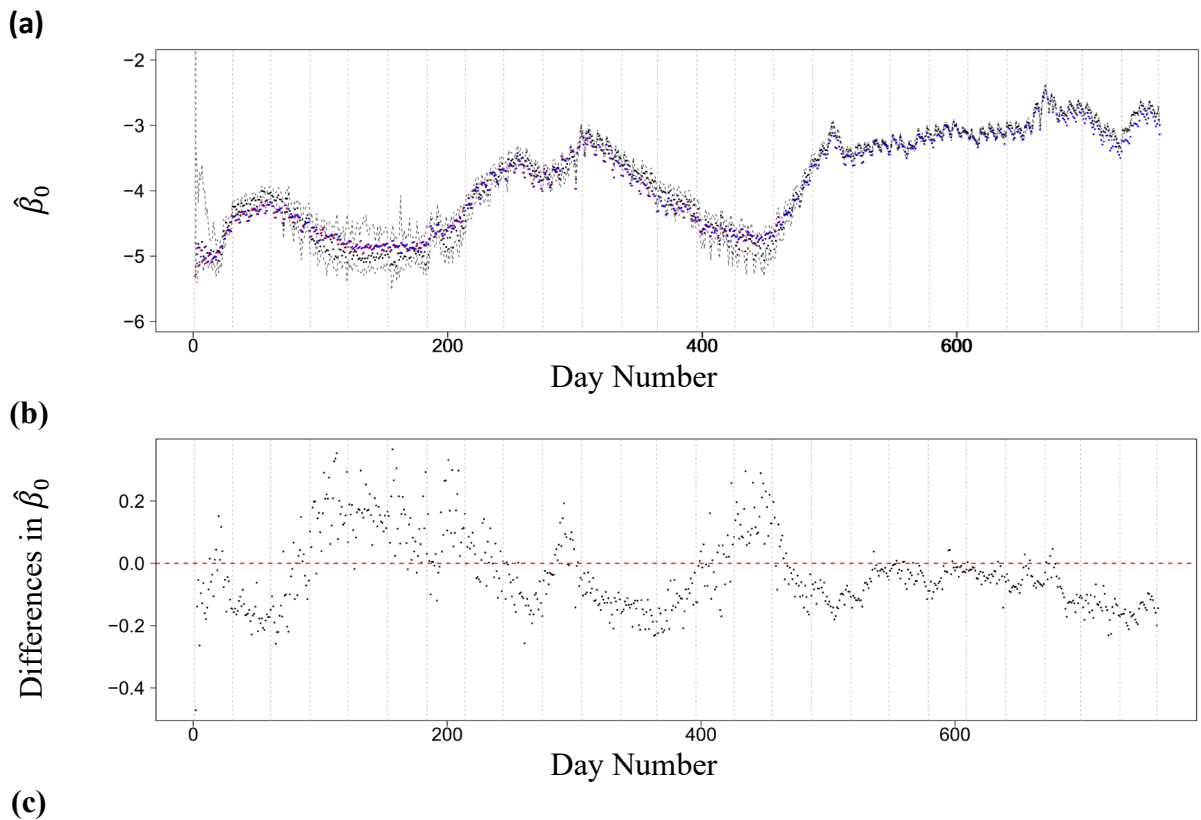

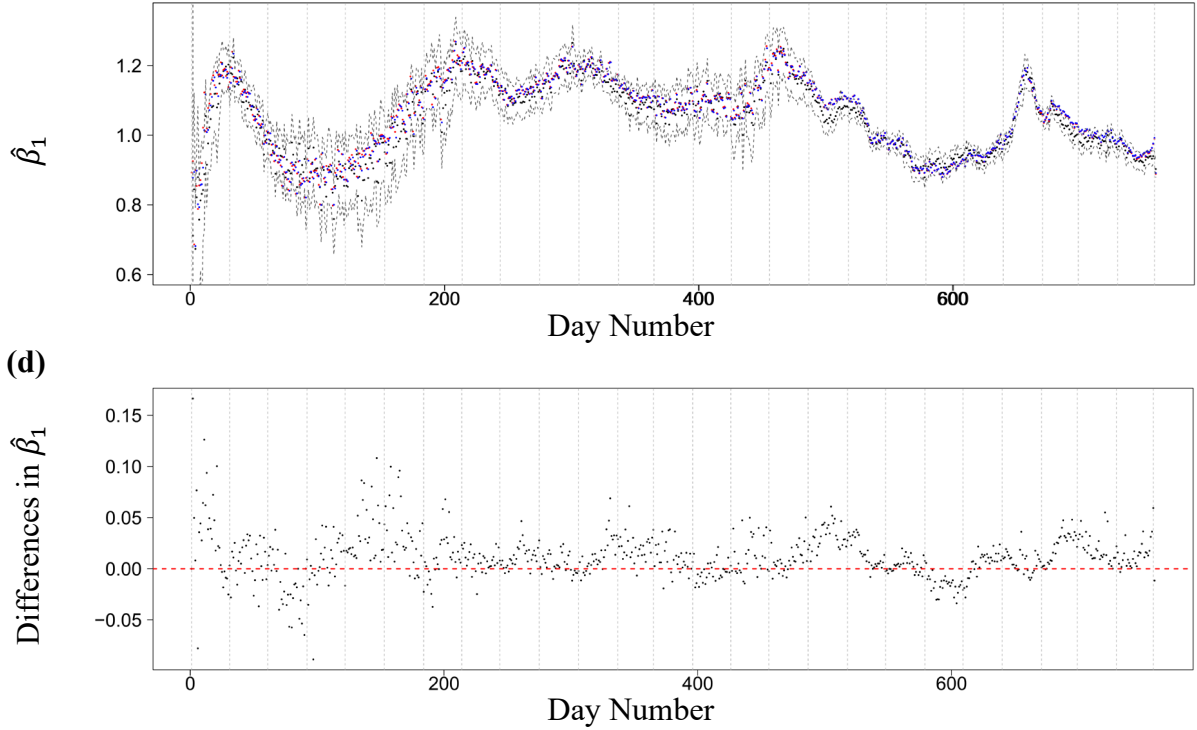

**(d)**

**Figure 5.** Daily time series of the estimated parameters and computed differences between models across 759 days. Panels show: PL pre-exponential factors $\beta_0$ (a), $\Delta\beta_0$ the difference between BGLR and BNR models (b), PL exponents $\beta_1$ (c), and $\Delta\beta_1$ (d) between BGLR and BNR models. In panels (a) and (c) blue dots represent the BNR model with varying variance, red dots represent the SLR model with fixed variance and the black dots represent the BGLR model. The grey dashed lines represent 95% credible intervals (with computed 2.5% and 97.5% quartiles) of the posterior distributions of the model parameters corresponding to the BGLR model. The horizontal dashed red line in panels (b) and (d) represents no differences in estimated parameters. A positive difference indicates a decreased BGLR parameter estimate while a negative difference indicates an increased BGLR parameter estimate.

A key advantage of the BGLR model is the variance regression coefficients accounting for the typical variance of GLM ($\beta'_0$) and scedasticity ($\beta'_1$) in the data (Figure 6). Normality methods usually assume a homoscedastic system. As shown previously [13] and further demonstrated here, the variance of the COVID data is complex. It was not constant and $\beta'_1$ (scedasticity) varied from -1 to 2.5. This additional information is indicated in the value of the $\beta'_1$ parameter. $\beta'_1 > 0$ indicates expanding variance with $x$ (Figure 6(d)). This was seen at the beginning of the pandemic, Summer 2020, December 2020, Spring 2021, October 2021 and December 2021 when rural regions (low population density) had lower variance. $\beta'_1 < 0$ indicates decreasing variance with $x$ (Figure 6(d)). This happened in April 2020, January-February 2021, July-August 2021, September 2021, November 2021 and January 2022 when urban areas had lower variance. This information is not provided in SLR models where it is assumed that variance is constant. The BNR model returned $\hat{\beta}'_1$ consistently close to 0 indicating homoscedastic variance throughout. The BNR model was biased since both the BGLR model (Figure 6c) and the data (Figure 6d and 6e) clearly indicate heteroscedasticity. The improved modelling of variance expands previous studies of COVID-19 [13] providing further information about the behaviour of population density on disease propagation.

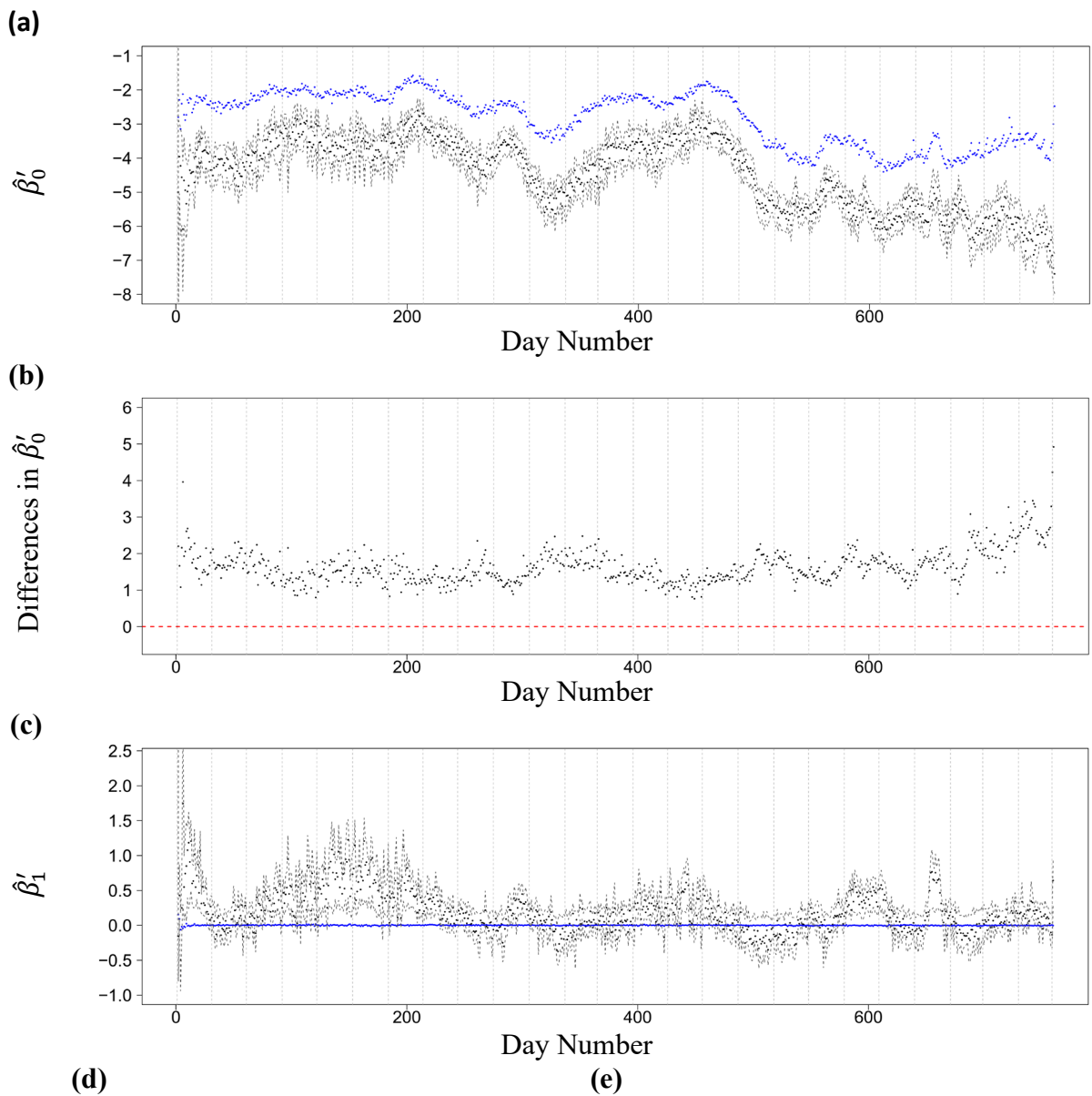

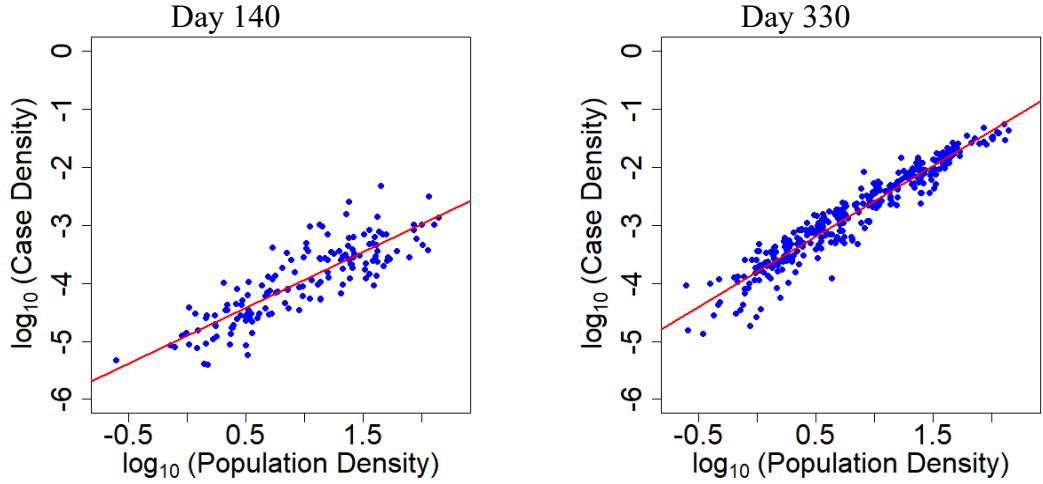

**Figure 6.** Daily time series of the estimated PL variance regression parameters along with computed differences across 759 days. Panels (a-b) are PL variance pre-exponential factors and differences between BGLR and BNR models. Panels (c), (d), and (e) illustrate the scedasticity parameter, $\hat{\beta}'_1$. In panels (a) and (c) blue dots represent the BNR model with varying variance and the black dots represent the BGLR model. The Grey dashed lines represent 95% credible intervals (with computed 2.5% and 97.5% quartiles) of the posterior distributions of the model parameters corresponding to the BGLR model. The horizontal dashed red line in (b) represents no differences in estimated parameters. A positive difference indicates a decreased BGLR parameter estimate while a negative difference indicates an increased BGLR parameter estimate. Positive and negative scedasticity are shown in panels (d) (day 140; $\hat{\beta}'_1 = 1.28 \ \{CI: lower = 1.05 \ upper = 1.46\}$) and (e) (day 330; $\hat{\beta}'_1 = -0.50 \ \{CI: lower = -0.58 \ upper = -0.39\}$).

The final parameter is the shape parameter, $\alpha$ (Figure 7a), allowing for skewed distributions. If $\alpha > 1$, the residual distribution is positively skewed (Figure 7(b)), if $\alpha = 1$ the residual distribution is symmetrical (Figure 7(c)) and for $0 < \alpha < 1$, the residual distribution is negatively skewed (Figure 7(d)). This data set was convincingly skewed over the period studied with periods dominated by both positive and negative skew. In this example, positively skewed residuals were indicative of 'hot spots' and 'super spreading' regions and negatively skewed residuals were indicative of 'cold spots' and 'super isolating' regions. The BGLR model reinforces the notion that skewed distributions are an unappreciated and recurring feature of disease propagation [13]. The BGLR model integrates skew within the model while rigorously accounting for heteroscedasticity.

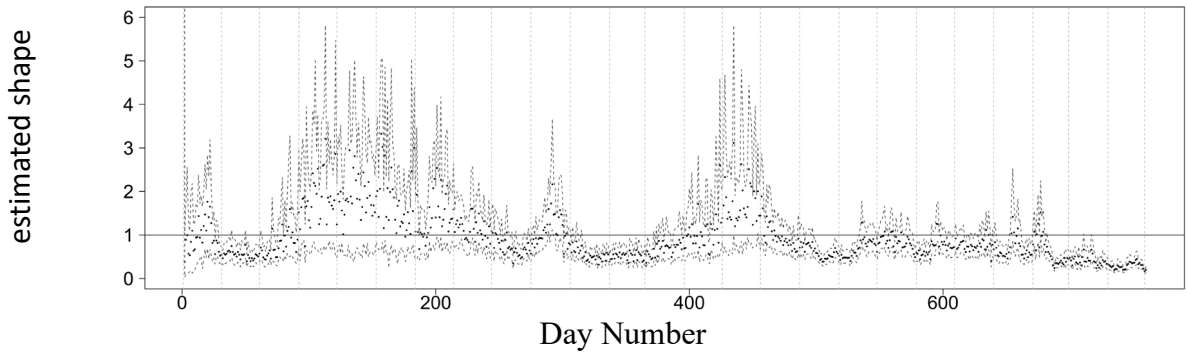

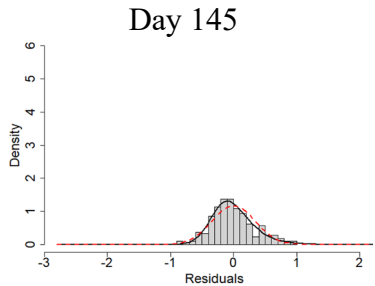 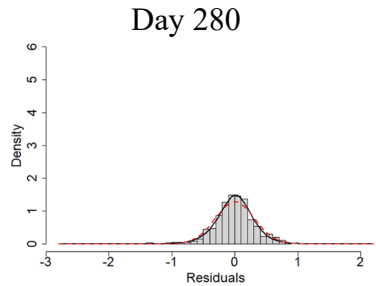 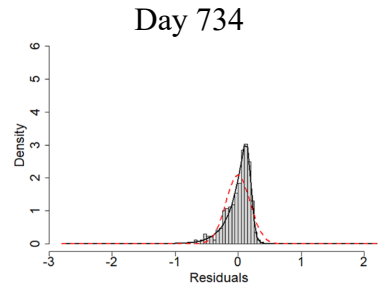

**Figure 7.** Daily time series of shape parameter and selected residual distributions. Panel (a) shows the time series of the shape parameter corresponding to the BGLR model (black dots). The Grey dashed lines represent 95% credible intervals (with computed 2.5% and 97.5% quartiles) of the posterior distributions of the model parameters corresponding to the BGLR model. The horizontal solid black line represents an unskewed (symmetric) distribution. Example residual distributions with (b) positive ($\hat{\alpha} = 1.98$ [$CI$:lower = 1.23, upper = 2.75]), (c) symmetrical ($\hat{\alpha} = 0.90$ [$CI$:lower = 0.66, upper = 1.20]) and (d) negative ($\hat{\alpha} = 0.19$ [$CI$:lower = 0.14, upper = 0.28]) skew in the residuals demonstrating the flexibility needed to model residuals throughout the pandemic. Density curves in panels c-d represent normal (dashed red line) and GL (solid black line) distributions fitted to residuals obtained from the BGLR model.

## 5.3 Model Comparison

To explore the preferred model, the DIC score was obtained for both Bayesian models (GL and normal) and differences were computed (Figure 8). A positive DIC corresponds to BGLR being the preferred model, whilst a negative value means BNR is preferred. The BGLR model is the preferred model for approximately 16 of 25 months in the data set. The periods of preference towards the Bayesian normal model, correspond to fewer regions reporting cases. For example, in the first 17 days of the studied period, there are fewer than 200 regions (out of 337 regions) reporting at least 1 case. Similar observations occur in summer 2020 and spring 2021. The BGLR model universally provided a better fit after day 480 (circa 23/06/2021), shortly after Delta spread. Nevertheless, our GL regression model allows for a symmetrical and asymmetrical response including positive and negative skew. This is not possible when exclusively using normality methods.

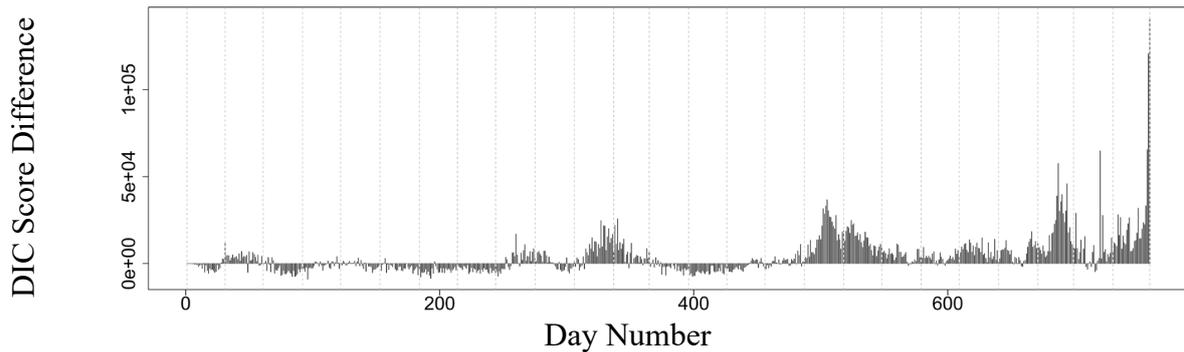

**Figure 8.** Daily time series of computed DIC difference scores between the BNR and BGLR models using Equation 18. A positive difference corresponds to the BGLR model as the preferred model. A negative difference indicates the BNR model is a better fit.

**6 Conclusions**

The introduction of the type I GLD within a GLM style framework provides key advances in regression methods for heteroscedastic and skewed data sets. We can now model heteroscedasticity with positively and negatively skewed data using a posterior distribution that may be analytically intractable. This was done in a Bayesian framework using MCMC methods. Specifically, the MH algorithm was applied with assumed normal regression coefficients with large variance (e.g. $10^4$) and an informative gamma shape parameter to express prior ignorance.

When applied to population density and COVID-19 case data the BGLR model tended to converge well throughout the pandemic and DIC scores indicated the BGLR model out performed normal methods during high observational timeframes. It demonstrated that a normal and homoscedastic model was insufficient to produce the complex behaviour exhibited during 759 days of the COVID-19 pandemic. It revealed varying skew and scedasticity which if ignored results in biased estimates of model parameters. These are previously unreported features of the COVID-19 data set. Although the application was COVID-19, this regression model could benefit any continuous data set with a sufficient number of points.

The BGLR uniquely models variance. The second moment corresponding to the GLD relates to a variance function including a shape parameter to account for skew. In our example we use the posterior median to confirm similar analysis of previous residual variance. However, adjusting the variance function to account for the linear predictor allows us to model variance of individual regions providing better local understanding of disease propagation. We find that periods with a high posterior median did not correspond to universal high variance across all regions. Instead, some timeframes exhibit a selection of regions that were the driving force of national heterogeneity.

This study has established a BGLR model that can characterize the full spectrum of important features of disease propagation. Previous studies of disease propagation have focused on distributions that only allow for positive skew. As previously documented [13] and further evident in this study, these models are inadequate to represent the complexity of infectious disease propagation over time. The BGLR models additional shape parameter exhibits skew ranging from strongly positive, indicative of 'hotspot' and 'superspreading' regions, to strongly negative, indicative of 'coldspots' and 'superisolating' regions. In addition, the BGLR model includes two additional variance parameters. The first of these parameters reveals extended periods of contraction indicative of regional homogeneity and expansion indicative of regional heterogeneity. The second variance parameter indicates heteroscedasticity, featuring both varying magnitudes of increasing and decreasing variance with population density. The BGLR model's ability to reveal these complexities demonstrate its utility for a host of scaling and other statistical problem.

Finally, the model incorporates a distribution (the GLD) where an analytic posterior distribution cannot be found or may not exist. The approach taken can be used as a template for other distributions to be used which have not yet been incorporated into a GLM style framework.

**Supplementary Material**

Supplementary material is available.

**Acknowledgements**

The authors are grateful to the UK Office of National Statistics and UK Health Security Agency for making these data available.

**Funding**

This research did not receive any specific grant from funding agencies in the public, commercial, or not-for-profit sectors.

**Disclosure Statement**

The authors report there are no completing interests to declare.

**Author contributions**

JS, GS, QH and HR contributed to the conceptualization, methodology, formal analysis and writing of original paper, reviewing and editing process. JS and QH assembled the data set.

**Supporting materials**

**Appendix A.** Data employed in this study. There are 337 lower tier local authorities reporting daily COVID-19 cases across 759 day. All data assembled in this study are included as supplementary information. This data was compiled from a range of publicly available sources as noted in the manuscript. They are provided as the following files: Regions.csv and COVID19Cases.csv.

**Appendix B.** All R code is available to replicate all the methods in this study along with the results obtained in the COVID-19 example. All R-scripts to assemble and analysis the data are included as supplementary information. This has been provided as BGLRModel.R, BNRModel.R, SNRModel.R, timeseriesResults.R, and GLDIllustrations.R.

**Appendix C.** Gelman and Rubin's convergence diagnostics.

**(a)**

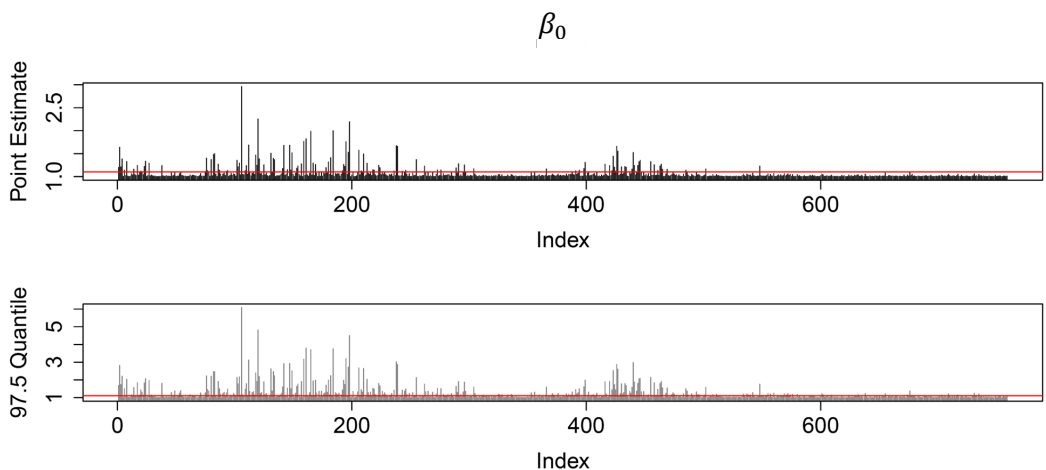

$\beta_0$

**(b)**

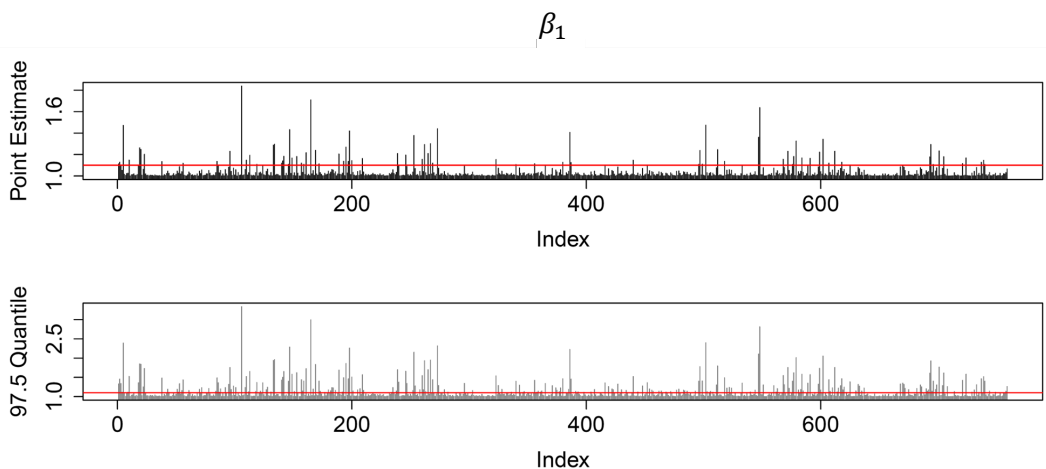

$\beta_1$

**(c)**

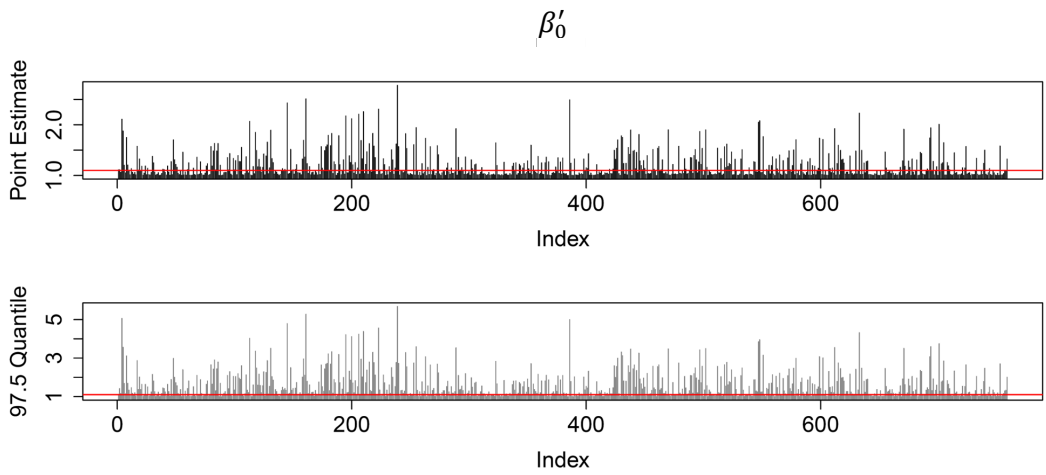

**(d)**

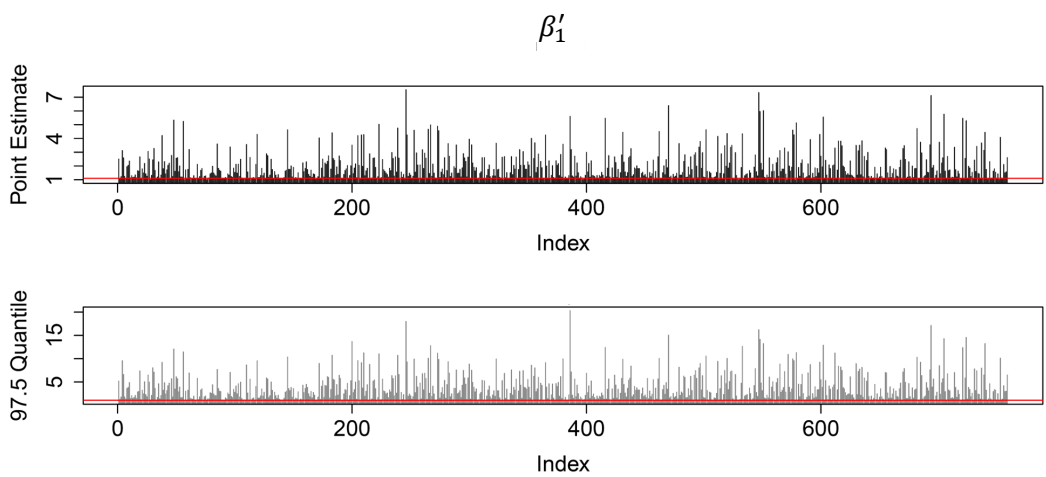

**(e)**

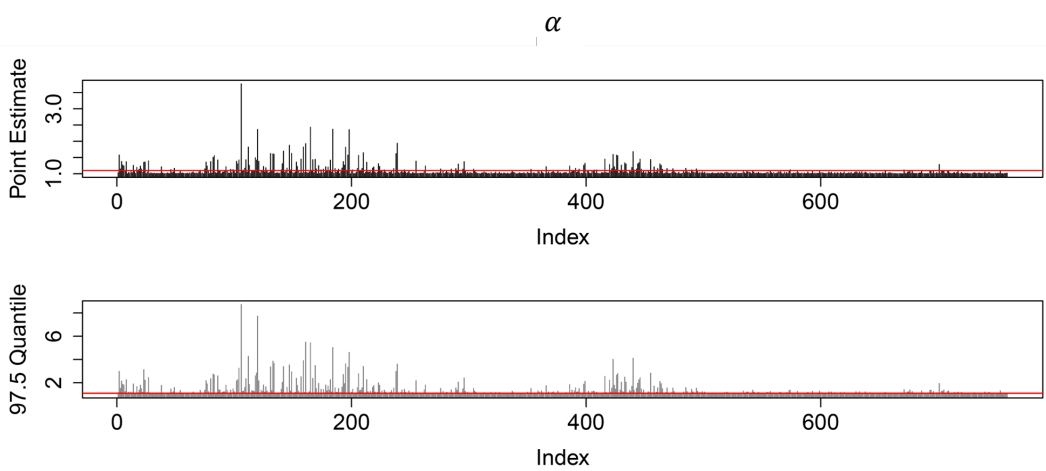

**Figure S1.** Daily Gelman and Rubin's R statistic over 20000 iterations for the parameter vector $\psi$ for the COVID-19 daily data set. Values less than 1.1 (represented by the horizonal solid red line) suggest that the MCMC chains converge.